# Application of integral equations to neutrino mass searches in beta decay

Thomas M. Semkow and Xin Li

*Wadsworth Center, New York State Department of Health, Albany, NY 12201, USA*
*and*
*School of Public Health, University at Albany, State University of New York,*
*Rensselaer, NY 12144, USA*

*Email: thomas.semkow@health.ny.gov*

## Abstract

A new mathematical method for elucidating neutrino mass from beta decay is studied. It is based upon the solutions of transformed Fredholm and Volterra integral equations. In principle, theoretical beta-particle spectra can consist of several neutrino-mass eigenvalues. Integration of the theoretical beta spectrum with a normalized instrumental response function results in the Fredholm integral equation of the first kind. This equation is transformed in such a way that the solution of it is a superposition of the Heaviside step-functions, one for each neutrino mass eigenvalue. A series expansion leading to matrix linear equations is then derived to solve the transformed Fredholm equation. Another approach is derived when the theoretical beta spectrum is obtained by a separate deconvolution of the observed spectrum. It is then proven that the transformed Fredholm equation reduces to the Abel integral equation. The Abel equation has a general integral solution, which is proven in this work by using a specific function for the beta spectrum. As an example, a numerical solution of the Abel integral equation is also provided, which has a fractional sensitivity of about $10^{-3}$ for subtle neutrino eigenvalue searches, and can distinguish from experimental beta-spectrum discrepancies, such as shape and energy nonlinearities.

# 1. Introduction

We start by outlining the physics of neutrino-mass searches. Since the neutrino was conceptually proposed by Pauli in 1930 and given its name by Fermi, followed by his development of the beta decay theory in 1933-1934 [1-3], studies of this elusive neutral particle have never ceased. Whether or not neutrino bears a rest mass involves the extension of the Standard Model, the composition of Dark Matter in the Universe, and the asymmetry between matter and antimatter [4]. An indirect evidence for neutrino mass has been provided by neutrino-oscillation experiments, which give us the differences between the masses of neutrinos with different flavors [5].

Three approaches are generally used to determine the absolute masses of neutrinos: investigation on the fluctuations of cosmic microwave background and large-scale structure, measurement of neutrino-less double-beta decay rate, and direct determinations of neutrino mass [6]. Currently, the first approach constraints the sum of three neutrino masses to less than 0.23 eV [7]. The second approach relies on the hypothesis that the neutrino is a Majorana particle, where a lower bound of the 0νββ decay half-life of $5.3\times10^{25}$ years has been set [8], whereas an effective neutrino mass of 0.1-0.9 eV was claimed [9].

The third approach has been focusing on a precise measurement of neutrino kinetic energy, as well as an endpoint region of a beta particle spectrum. The former requires a very strong neutrino source: the analysis of the neutrino energy spectrum as a function of arrival time from SN1987A supernova data yielded an upper limit of neutrino mass at 5.7 eV [10]. In case of the latter, the Mainz and Troitsk experiments for $^3$H beta decay yielded the upper limits of electron antineutrino mass at 2.3 eV [11] and 2.05 eV [12], respectively, while the KATRIN experiment is expected to have a sensitivity down to 0.2 eV [13]. The $^{187}$Re beta-decay bolometric experiments resulted in upper limits for neutrino mass of 26 eV [14] and 15 eV [15]. The $^{163}$Ho (electron capture) neutrino experiments yielded higher upper limits of 225 eV [16] and 490 eV [17] and efforts are in progress to lower these limits in the ECHO experiment [6].

In the science of direct search for massive neutrino in beta decay away from the endpoint energy, there were claims of the discovery of 17-keV neutrino in $^3$H [18], $^{14}$C [19], and $^{35}$S [20,21]. The 17-keV component appeared to be present as a discontinuity in beta spectra at a fractional contribution of approximately 0.01. These effects were later explained as electrons scattering from collimators [22,23].

The remainder of the paper concentrates on the neutrino mass searches in β⁻ decay (hereinafter referred to as beta decay). The electron antineutrino, $\bar{\nu}_e$ (hereinafter referred to as neutrino) is emitted in the weak-interaction process of neutron decay inside the nucleus: n ⟶ p⁺ + β⁻ + $\bar{\nu}_e$. The emitted β-particle (electron) kinetic-energy spectrum is a continuous function, owing to the energy sharing between the beta particle and the neutrino. Possible neutrino mass $M$ in the beta spectrum is included in the factor: $\sqrt{(Q-T)^2 - M^2}$, where $Q$ is the nuclear-recoil corrected $Q$-value of beta decay and $T$ is the emitted electron kinetic energy. The maximum beta energy (called the endpoint energy) is equal to $Q$ or $Q - M$, when the neutrino mass is zero or positive, respectively. In principle, there can be several mixed neutrino eigenvalues. The details of the theoretical beta spectra are described in Section 2.

The emitted beta spectrum is convoluted with the instrumental response to yield the observed beta spectrum. The existing methods of identification of the neutrino mass in beta decay, whether close to or away from the endpoint energy, employ two approaches: i) convolution of the theoretical beta spectrum with the normalized instrumental response function and comparison of the convolution with the experimental spectrum, and ii) deconvolution of the observed spectrum and comparison with the theoretical spectrum. Statistical measures, such as $\chi^2$ minimization or Bayesian likelihood, are used for the above-mentioned comparisons. Statistical measures can be non-specific, *i.e.*, there may be several factors other than the neutrino mass, which affect them.

Therefore, we aimed in this work at developing mathematical method based on transformational properties, rather than statistical measures, to elucidate neutrino mass in beta decay. In Section 3, we derive a transformed Fredholm equation of the first kind, which includes both theoretical beta spectrum and instrumental response function, and show that the solution to it is a superposition of Heaviside step-functions with an abscissa of $(Q-T)^2$, one for each possible neutrino mass eigenvalue. The neutrino mass can then be identified by the abscissa value at the raise of the step function, whereas its eigenvalue contribution is shown to be proportional to the ordinate value past the step. In Section 4, we outline a possible solution to the transformed Fredholm equation using series expansion and linear equations in a matrix notation.

If method ii) of deconvoluting the beta spectrum first, before comparison with the theory, is employed, then the transformed Fredholm equation from Section 3 is shown in Section 5 to reduce to the Abel integral equation, also known as the Abel transformation. The Abel integral equation has a known integral solution. We prove in Section 5 that, if the function undergoing Abel transformation has a form proportional to the beta spectrum, then the solution to the Abel equation is a Heaviside step-function. In Section 6, we provide a numerical example of a solution of the Abel equation, when the beta spectrum contains several neutrino-mass eigenvalues. We also describe in detail how the Abel solution behaves in the presence of small experimental discrepancies, which can influence neutrino mass detection, such as β shape correction or small nonlinearities in the measured β-energy scale. Finally, in Section 7 we summarize the advantages and limitations of the proposed new mathematical method of elucidating neutrino mass in beta decay, and propose future directions of this work.

## 2. Beta-decay theory

There exist established formalisms for calculating allowed and superallowed beta spectra, which differ in some approaches [6,24,25]. For the purpose of this work, we define the beta-energy spectrum as follows:

$$\dot{N}(T) = AD(T)(Q-T)\sqrt{(Q-T)^2 - M^2}, \qquad (1)$$

where $\dot{N}(T) = dN/dT$ represents a distribution of beta particles with energies between and $T$ and $T + dT$, factor $A$ contains the quantum-transition matrix element and an overall normalization, whereas $Q, T,$ and $M$ have been defined in Section 1.

Factor $D(T)$ is defined as follows:

$$D(T) = pEF_n F_r C_r S, \qquad (2)$$

where $p$ is relativistic momentum, $E$ is total (relativistic) energy, $F_n$ is a non-relativistic Fermi function, $F_r$ is a relativistic correction to the Fermi function [26], and $C_r$ is a radiative correction [24]. $S$ is a shape correction [24], which corrects for small smooth discrepancies between the theoretical and measured spectra at low energies, is assumed to be a parabola with an argument of $Q - T$. Also, a screening correction [27,28] is applied to $D(T)$. Additional terms involving the daughter's final excitation states in low beta-energy emitters [6], as well as a possible Lorentz-invariance violation correction [29], have not been included.

The examples of beta spectra from $^3$H decay ($Q = 18590.29$ eV) [30] calculated according to Eqs. (1,2) are plotted in Fig. 1. The physical constants and nuclear parameters for the calculations were taken from Ref. [30]. It is seen that if neutrino mass is non-zero, the endpoint beta energy lies below $Q$, while another neutrino mass eigenvalue is evidenced by the discontinuity of the first derivative in the spectrum. If neutrino mass is close to zero and any other eigenvalue has a very small contribution, these effects are difficult to discern from the beta spectrum.

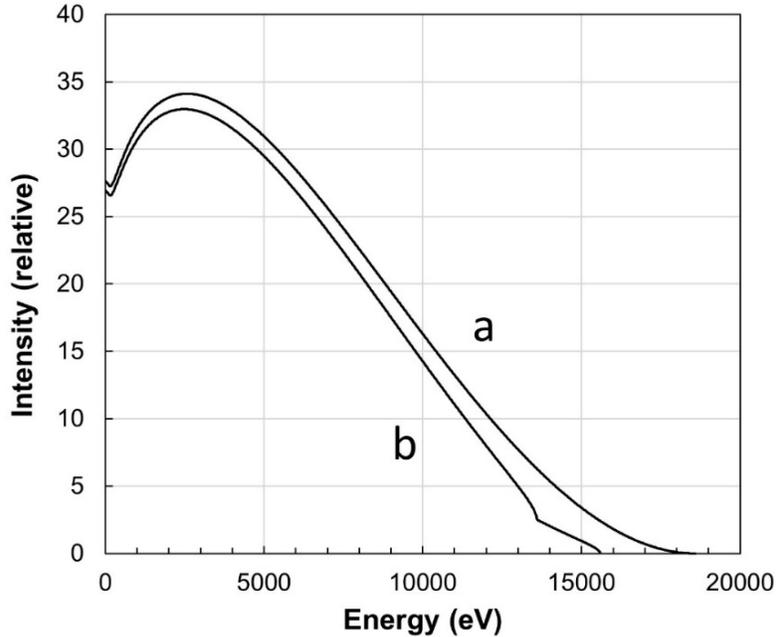

Figure 1. Beta particle spectra from $^3$H decay. Spectrum (a): zero neutrino mass; spectrum (b): equal superposition of two neutrino eigenvalues with the masses of 3000 and 5000 eV.

## 3. Transformed Fredholm integral equation

The observed beta spectrum usually differs from the emitted (theoretical) beta spectrum, given by Eqs. (1,2), owing to an instrumental response. Let $T$ represent the emitted kinetic energy while $T'$, the observed kinetic energy. The emitted and observed beta spectra are then $\dot{N}(T)$ and $\dot{N}(T')$, respectively. They are coupled with each other by the response function $R(T',T)$ through the integral:

$$\dot{N}(T') = \int_0^{Q-M} R(T',T)\dot{N}(T)dT. \qquad (3)$$

Equation (3) is a Fredholm integral equation of the first kind [31], where $R(T',T)$ is a kernel. The variable $T'$ can extend past $Q - M$ due to the abovementioned instrumental response. The response function is usually measured separately. Since the emission and detection of beta particles are stochastic processes, the normalized $R(T',T)$ and $\dot{N}(T)$ can be considered statistical probability density functions (pdf). The integral in Eq. (3) is then a mixture of distributions [32]. If, additionally, $R(T',T)$ is a function of $T' - T$, e.g., a Gaussian, then Eq. (3) is statistically a convolution [33]. The process of solving for $\dot{N}(T)$ is referred to as deconvolution and is discussed in Section 4. The terms: convolution and deconvolution will be used here even if Eq. (3) is not a strictly mathematical convolution.

If there existed $k$ neutrino-mass eigenstates with the masses $M_k$ and normalization factors $A_k$, then combining Eqs. (1,3) and summing over the eigenstates would result in the observed spectrum:

$$\dot{N}(T') = \sum_k A_k \int_0^{Q-M_k} R(T',T)D(T)(Q-T)\sqrt{(Q-T)^2 - M_k^2}dT. \qquad (4)$$

Subsequently, we introduce a mass variable $m$ and utilize the Dirac $\delta$-function [34]. Then Eq. (4) transforms to:

$$\dot{N}(T') = \sum_k A_k \int_0^Q dm\, \delta(m - M_k) \int_0^{Q-m} dT\, R(T',T)D(T)(Q-T)\sqrt{(Q-T)^2 - m^2}. \qquad (5)$$

New variables are defined: $x = (Q - T')^2$ and $z = (Q - T)^2$. By substitution, Eq. (5) results in

$$\dot{N}(x) = 1/2 \sum_k A_k \int_0^Q dm\, \delta(m - M_k) \int_{m^2}^{Q^2} dz\, R(x,z)D(z)\sqrt{z - m^2}. \qquad (6)$$

By using a property of the $\delta$-function [34],

$$\delta(m - M_k) = 2m\, \delta(m^2 - M_k^2), \qquad (7)$$

and substituting of the variables, $y = m^2$, $a = Q^2$, and $a_k = M_k^2$, one obtains from Eq. (6):

$$\dot{N}(x) = 1/2 \sum_k A_k \int_0^a dy\, \delta(y - a_k) \int_y^a dz\, R(x,z)D(z)\sqrt{z - y}. \qquad (8)$$

Changing the order of integration in Eq. (8), yields

$$\dot{N}(x) = 1/2 \sum_k A_k \int_0^a dz\, R(x,z)D(z) \int_0^z dy\, \delta(y - a_k)\sqrt{z - y}. \qquad (9)$$

The inner integral in Eq. (9) can be integrated by parts. We also use the fact that the integral of the $\delta$-function on positive argument is the Heaviside step function:

$$\int \delta(y - a_k)\, dy = H_{a_k}(y) = \begin{cases} 0, & y \leq a_k \\ 1, & y > a_k \end{cases}. \tag{10}$$

The result is given below:

$$\dot{N}(x) = 1/4 \sum_k A_k \int_0^a dz\, R(x,z) D(z) \int_0^z dy\, H_{a_k}(y)/\sqrt{z-y}. \tag{11}$$

The solution to Eq. (11) is the function

$$f(y) = \sum_k A_k H_{a_k}(y), \tag{12}$$

which is a superposition of the step functions for all possible neutrino mass eigenvalues, scaled by their appropriate contributions. By inserting (12) into (11) one obtains

$$\dot{N}(x) = 1/4 \int_0^a dz\, R(x,z) D(z) \int_0^z dy\, f(y)/\sqrt{z-y}. \tag{13}$$

By reversing the order of integration in Eq. (13) back to that of Eq. (8) and by dividing both sides by $D(x)$, i.e., the theoretical factor $D$ given by Eq. (2) evaluated at the observed energy $T'$, one obtains

$$\frac{\dot{N}(x)}{D(x)} = \int_0^a K(x,y) f(y) dy, \quad \text{with} \tag{14a}$$

$$K(x,y) = 1/4 \int_y^a \frac{R(x,z)}{\sqrt{z-y}} \frac{D(z)}{D(x)} dz. \tag{14b}$$

Equation (14a) is the transformed Fredholm equation. The kernel is given by Eq. (14b). It links the observed beta spectrum with the emitted (theoretical) spectrum and with the instrumental response. In the solution function given by Eq. (12), one does not need to assume the presence of neutrino mass or its eigenvalues. They may be revealed by the abscissa and ordinate of any step functions present in the solution, to within the accuracy of the solution, which is dependent on the detailed shape of the observed spectrum. Therefore, the proposed approach interrogates the shape of the observed spectrum using the transformation derived, in contrast to the methods based on statistical measures.

## 4. Solution of the Fredholm integral equation

There exist algorithms for solving the Fredholm equations given by Eqs. (3) and (14a) [35,36]. If Eq. (3) is a mathematical convolution, then it can also be solved by the Laplace transform method, taking the advantage of the theorem that the Laplace transform of the convolution is a product of Laplace transforms of the convoluting functions [37]. However, Eq. (14a) is not a convolution and we derive below a matrix method to solve it.

We take into consideration the fact that the observed $T'$ variable is discrete and incremented at equally spaced intervals (channels) owing to the instrumental digitizing of the spectrum (modern instrumentation can have $n = 2^{15}$ or more channels). Consequently, variable $x$ is also digitized. Consider a general integral equation of the first kind, of which Eqs. (3,14a) are the examples:

$$\phi(x) = \int K(x, y)\psi(y)\, dy. \tag{15}$$

If $x$ is discrete, one has:

$$\phi(x_i) = \int K(x_i, y)\psi(y)\, dy, \qquad i = 1, \ldots, n. \tag{16}$$

Let us assume a trial solution in a form of a series:

$$\psi(y) = \sum_{j=1}^{n} K(x_j, y)\, c_j. \tag{17}$$

By inserting the trial solution into Eq. (16), one obtains a set of linear equations:

$$\phi(x_i) = \sum_{j=1}^{n} K_{ij}\, c_j, \qquad i = 1, \ldots, n, \tag{18a}$$

with

$$K_{ij} = \int K(x_i, y) K(x_j, y)\, dy. \tag{18b}$$

In matrix notation, Eq. (18a) reads:

$$\mathbf{\Phi} = \mathbf{KC}, \tag{19}$$

where $\mathbf{\Phi}$ and $\mathbf{C}$ are vectors and $\mathbf{K}$ is a symmetric square matrix. By solving the linear equations for $\mathbf{C}$, the solution (17), continuous in $y$ variable (i.e., in $(Q - T)^2$) can, in principle, be obtained.

## 5. Abel integral equation and its solution

In Section 3, the general treatment for elucidating of neutrino mass in beta decay was presented, leading to the transformed Fredholm equation (14a). This equation can be considerably simplified if the observed beta spectrum is first deconvoluted using Eq. (3). The limit to the transformed Fredholm equation (14a) can then be obtained by using the fact that, if the spectrum was deconvoluted, then for the purpose of (14a),

$$R(T', T) = \delta(T' - T) = \delta[(Q - T) - (Q - T')] = \delta(\sqrt{z} - \sqrt{x}) = 2\sqrt{z}\delta(z - x),$$

where we used a version of Eq. (7). Then, starting from Eq. (13),

$$\dot{N}(x) = 1/4 \int_0^a dz\, \delta(z - x) 2\sqrt{z} D(z) \int_0^z dy\, f(y)/\sqrt{z - y}$$

$$= 1/2\, D(x)\sqrt{x} \int_0^x f(y)/\sqrt{x - y}\, dy. \tag{20}$$

By defining a new function:

$$g(x) = \frac{\dot{N}(x)}{D(x)\sqrt{x}}, \tag{21}$$

we obtain from Eq. (20)

$$g(x) = \frac{1}{2} \int_0^x \frac{f(y)}{\sqrt{x - y}}\, dy. \tag{22}$$

Equation (22) is a special case of the Abel transformation, also called the Abel integral equation [31,38], which is a special case of the Volterra integral equation of the first kind. The general form of the Abel equation is

$$g(x) = \int_0^x (x-y)^{-\alpha} f(y)\, dy, \qquad 0 < \alpha < 1, \tag{23}$$

with a solution of

$$f(x) = \frac{\sin \pi\alpha}{\pi} \frac{d}{dx} \int_0^x (x-y)^{\alpha-1} g(y)\, dy. \tag{24}$$

Therefore, the solution of Eq. (22) for $\alpha = 1/2$ is

$$f(x) = \frac{2}{\pi} \frac{d}{dx} \int_0^x \frac{g(y)}{\sqrt{x-y}}\, dy. \tag{25}$$

By using Eq. (1) for a single neutrino-mass eigenvalue $M$, and setting $y = (Q-T)^2$ as well as $b = M^2$, we obtain from Eq. (21)

$$g(y) = A\sqrt{y-b}. \tag{26}$$

By inserting Eq. (26) to Eq. (25), one obtains

$$f(x) = \begin{cases} 0, & y \leq b \\ \dfrac{2A}{\pi} \dfrac{d}{dx} \displaystyle\int_b^x \dfrac{\sqrt{y-b}}{\sqrt{x-y}}\, dy, & x > y > b. \end{cases}$$

$$\int_b^x \frac{\sqrt{y-b}}{\sqrt{x-y}}\, dy \xrightarrow{\text{by parts}} \frac{1}{2}(x-b)\int_b^x \frac{dy}{\sqrt{(y-b)(x-y)}} = \frac{\pi}{2}(x-b),$$

since the last integral is equal to $\pi$ [39]. Whence,

$$f(x) = \begin{cases} 0, & x \leq b \\ A\dfrac{d}{dx}(x-b) = A, & x > b. \end{cases} = AH_b(x).$$

Therefore, the solution in Eq. (25) is a Heaviside step function for the beta spectrum.

If there are $k$ neutrino mass eigenstates $M_k$, $a_k = M_k^2$, with relative contributions $A_k$, then the function $g$ from Eq. (26) can be written as:

$$g(y) = \sum_k A_k \sqrt{y - a_k}, \tag{27}$$

and a generalization of the solution to the Abel equation is given by Eq. (12).

## 6. Numerical examples

In this Section, several numerical examples are provided for the solution to the Abel equation. In a possible applications of this method to the experimental spectrum, test function $g$ is defined as the ratio of the deconvoluted β-energy spectrum $\dot{N}(T)$, obtained from Eq. (3), to calculated $D(T)$. The function $g$ is given by Eq. (21). Since we do not know a priori how many eigenvalues there are and what their corresponding neutrino masses are, one needs to perform a numerical solution on test function $g$ using Eq. (25). However, for the purpose of this demonstration, we assume one or more components in a simulated spectrum $\dot{N}(T)$, calculated according to Eq. (1), without deconvolution. For numerical integration and differentiation of the test function using Eq. (25), we apply the established high-performance algorithms [35] with an additional adaptive optimization of the differential step.

In the first example, we demonstrate how the step-function solution $f$ depends on the bin size of abscissa $x$, for mass-less neutrino. Several step functions for zero neutrino-mass $^3$H decay ($Q$ = 18590.29 eV) [30] are plotted in Fig. 2 as functions of the $x = (Q - T)^2$ argument for different bin sizes. The leftmost points at $x = 0$ could not be plotted on a logarithmic scale, so they were placed on the abscissa at the smallest bin studied, $x = 0.01$ eV$^2$. Therefore, the position of the step for a given bin size can be taken as an upper limit of the square of neutrino mass.

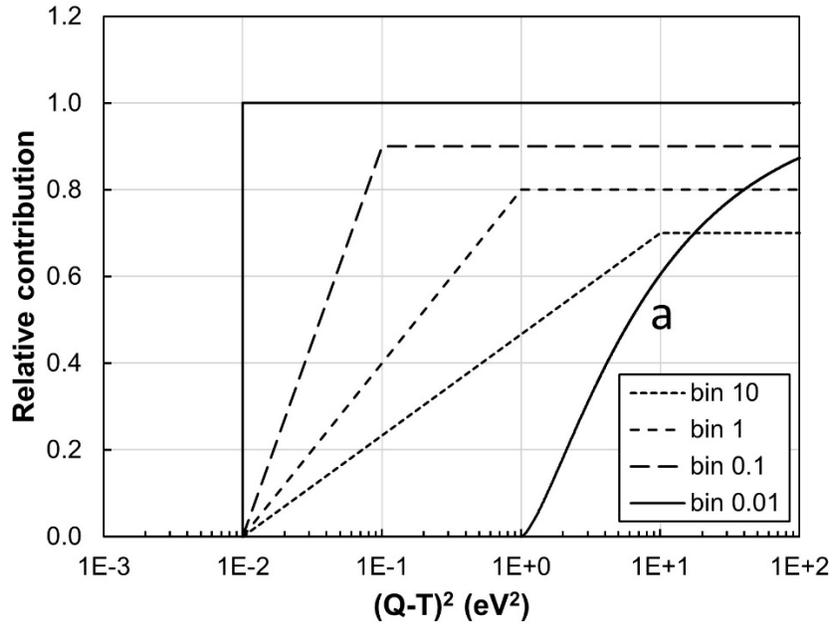

Figure 2. Numerical solutions of the Abel equation for zero neutrino mass $^3$H decay (except curve (a)), plotted for various abscissa bin sizes in eV$^2$ indicated in the legend. Zero-abscissa values are arbitrarily placed at 0.01 eV$^2$. The relative contributions are scaled to 1.0, 0.9, 0.8, and 0.7. Curve (a) depicts the effect of energy nonlinearity for 0.01 eV$^2$ bin.

In the second example, we selected beta decay of $^{35}$S ($Q$ = 167.319 keV [30]) and assumed 3 neutrino mass eigenvalues at 0, 10, and 15 keV as well as their contributions (normalization factors) as given in Table 1. Function $g$ was calculated using Eq. (27), and subjected to the numerical solution using Eq. (25). The solution $f$ is plotted as a function of argument $(Q - T)^2$ in Fig. 3 (solid curve), emphasizing the vicinity of 1 on the ordinate axis. It is seen that the three step functions are easily identified. The raise of the step functions on the abscissa of the numerical solution correctly identify squares of the neutrino masses at 0, 100, and 225 keV, respectively. The identified contributions agree with the assumed contributions. While the method is in principle non-statistical, the relative uncertainties of the identified contributions are caused by rounding errors in the numerical solution. The relative uncertainty is negligible for the principal eigenvalue and between 1-2% for the minor eigenvalues.

Table 1. Assumed and identified parameters for 3-neutrino mass eigenvalue test of the numerical solution of the Abel equation for searching of neutrino mass in beta decay of $^{35}$S.

| Parameter | Eigenvalue | | |
| --- | --- | --- | --- |
| | 1 | 2 | 3 |
| Assumed neutrino mass (keV) | 0 | 10 | 15 |
| Assumed contribution | 1 | 0.001 | 0.001 |
| Identified contribution | 1.000E+00 | 9.997E-04 | 1.000E-03 |
| Relative uncertainty | 2.2E-11 | 1.2E-02 | 1.7E-02 |

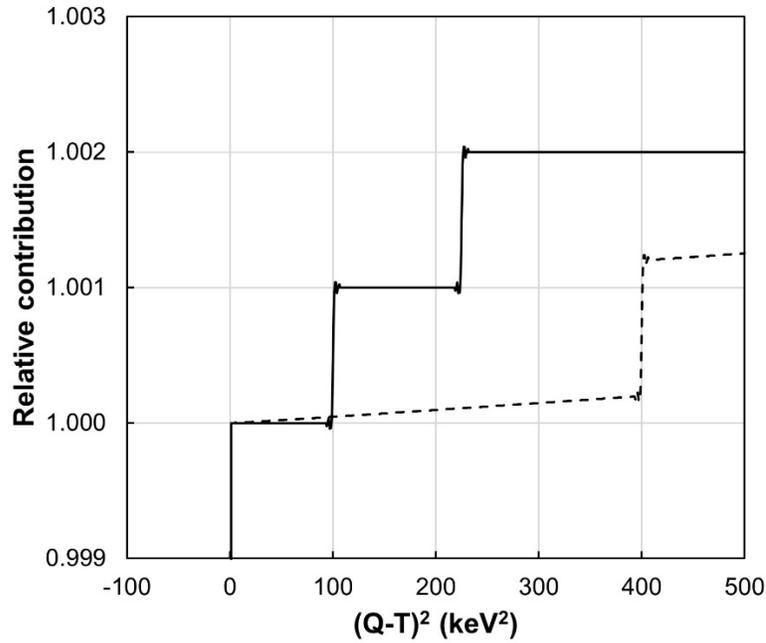

Figure 3. Numerical solution of the Abel equation for searching of neutrino mass in the beta decay of $^{35}$S. Solid curve: three neutrino components with masses of 0, 10, and 15 keV. Dashed curve: two neutrino components with masses 0 and 20 keV with a shape correction function.

When comparing experimental and theoretical β spectra, small smooth discrepancies were observed, which increase with decreasing kinetic energy [24]. They can be corrected for with a shape-correction factor $S$, which is assumed to be a parabola with an argument $Q - T$. An example of such a shape correction function is given in Fig. 4 (solid curve) for $^{35}$S decay. This correction is equal to zero at a maximum kinetic energy, and it is assumed 1.01 at zero energy. In the following, we analyzed the effect of this correction on the Abel solution. We assumed that $\dot{N}(T)$ had a deviation from a theoretical shape according to the shape correction function in Fig. 4, included in $D(T)$ in Eq. (2), and did not correct for it when dividing by $D(T)$ in Eq. (21). We also assumed two neutrino mass eigenstates at 0 (contribution of 1) and 20 keV (contribution of 0.001). Subsequent application of the Abel solution resulted in a dashed curve in Fig. 3. It is seen that not applying shape correction results in the flat portions of the step function sloping upward, however, both the neutrino mass and its contribution were unaffected.

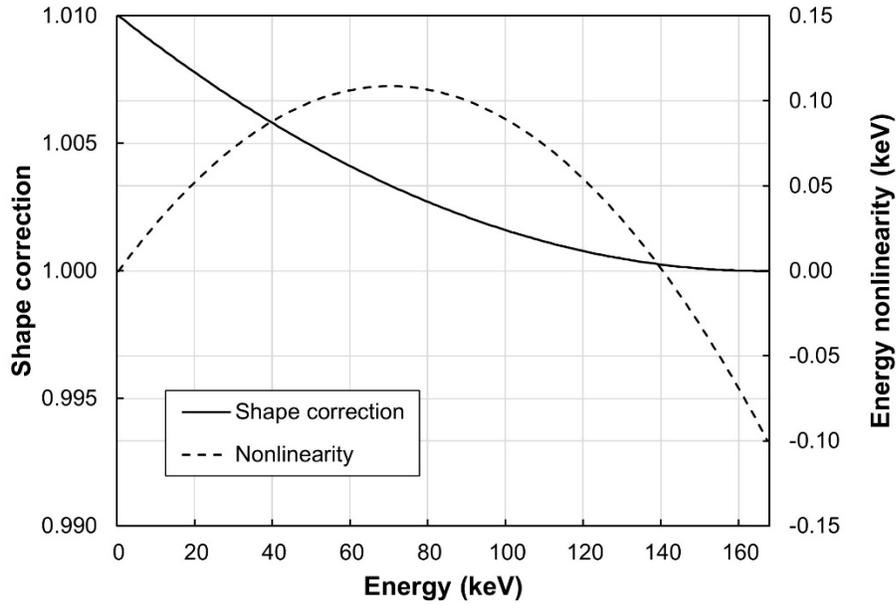

Figure 4. Shape correction and energy nonlinearity plotted as functions of energy for $^{35}$S decay.

Besides the shape correction, which affects the intensity of the β spectrum (in the ordinate direction), we study the effect of energy-scale nonlinearities (in the abscissa direction) on the Abel solution and neutrino-mass detection for $^{35}$S decay. Such minute nonlinearities may result from improper energy calibration or from nonlinear response of a digital signal processor. As a next example, we assumed the energy nonlinearity with functional dependence as depicted in Fig. 4 (dashed curve). We calculated $\dot{N}(T)$ with the energy nonlinearities included and did not correct for them in $D(T)$ for Eq. (21). We study the effects close to the maximum kinetic energy (endpoint energy, at zero neutrino mass) and at lower kinetic energies (for a possible presence of heavy neutrino) separately.

For any nonlinearity that is positive at the end-point energy and is not corrected for, the measured energy would exceed $Q$ and thus would be identified as an experimental problem. Therefore, we assumed nonlinearity equal to –0.1 keV at the maximum kinetic energy (see Fig. 4). Such nonlinearity would result in the measured energy being shifted down by 0.1 keV (~ 0.06%) below $Q$. This might simulate the presence of neutrino mass. The Abel solution for this case is depicted in Fig. 5. Without nonlinearities present (dashed curve), the upper limit on the square of neutrino mass would be placed at 0.001 keV$^2$. The above-mentioned nonlinearity appear to simulate a square of the neutrino mass at 0.01 keV$^2$ (solid curve), however, it can be rejected because the step is not sharp but is a rather slowly raising function instead. The effect of the energy nonlinearity from Fig. 4 on the wider range of the Abel solution is depicted in Fig. 6 (solid curve), where we have also added a 100-keV neutrino component at 0.001 contribution. It is seen that the curve resembles an exact step, as it does in the case without nonlinearities depicted by the dashed curve. Therefore the heavy neutrino is identified at 100 keV and only its mass is shifted by 0.1 keV nonlinearity (see Fig. 4).

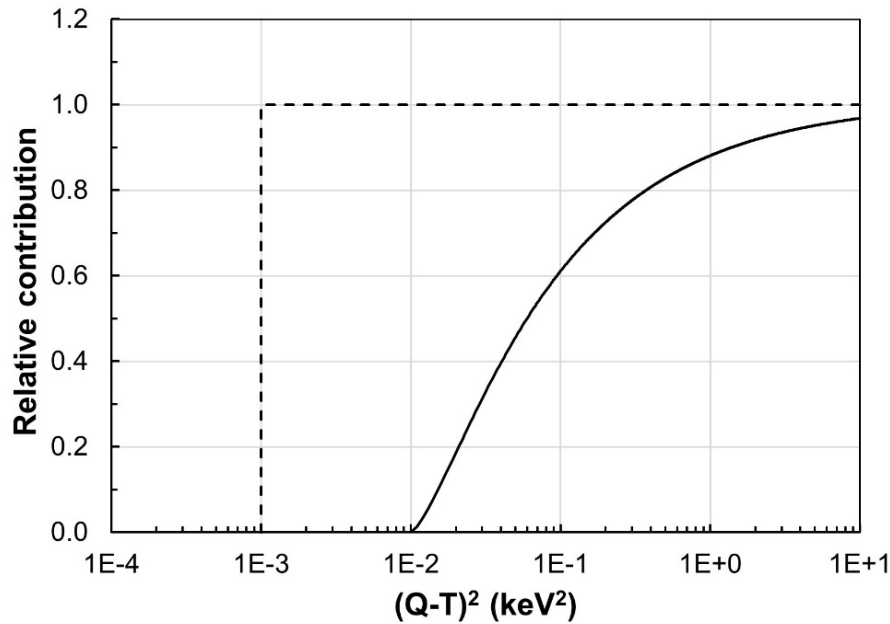

Figure 5. Step function solutions for zero neutrino mass $^{35}$S decay, with (solid curve) and without (dashed curve) energy nonlinearities.

We have also assessed the effect of energy nonlinearities on neutrino mass in $^3$H decay. We assumed a functional nonlinearity similar to the one in Fig. 4, except the energy shift at the endpoint was assumed to be –1 eV (~ 0.005%) below $Q$. The Abel solution resulted in curve (a) in Fig. 2. It appears to simulate a square of the neutrino mass at 1 eV$^2$, however, it can be rejected following the arguments for $^{35}$S in Fig. 5.

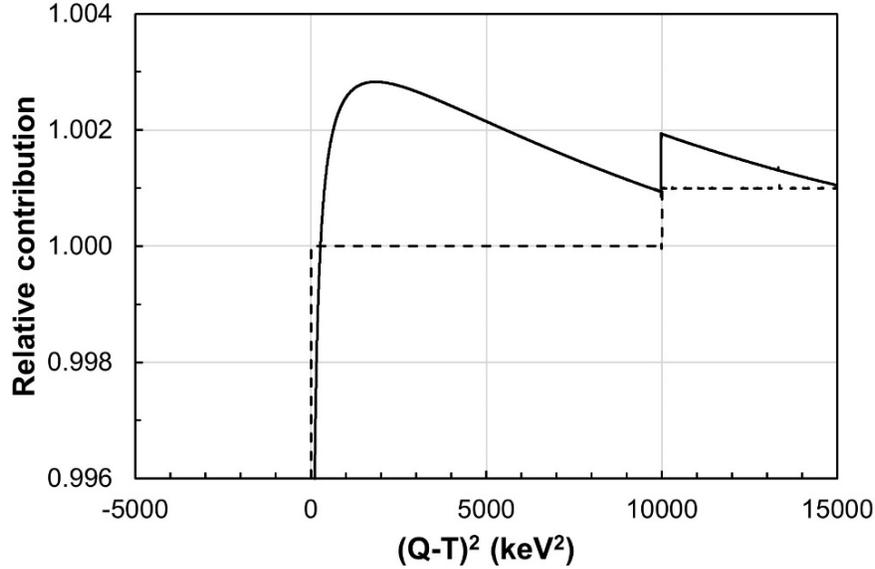

Figure 6. Step function solutions for $^{35}$S decay consisting of two neutrino components having masses of 0 and 100 keV, with (solid curve) and without (dashed curve) energy nonlinearities.

## 7. Summary and conclusions

We have developed a math-theoretical model for searches of neutrino mass in beta-decay energy spectra. The transformed Fredholm equation was derived (Eq. (14a) with the kernel given by Eq. (14b)), linking the observed beta spectrum with the theoretical one as well as with the instrumental response. The solution to this equation is a superposition of the Heaviside step-functions representing the possible neutrino mass eigenvalues, as well as their contributions (Eq. (12)). In this way, the shape of beta spectrum is tested for discrepancies, which could be associated with the neutrino mass. The step function is a very characteristic signature of such discrepancies. A possible solution to the Fredholm equation was derived based on series expansion (Eq. (17)) and solution of the matrix equation (Eq. (19)), taking advantage of the fact that the observed spectrum is discrete. Numerical verification of the transformed Fredholm equation deserves further investigation.

A limiting version of the transformed Fredholm equation was also derived. It is based upon regular deconvoluting of the observed spectrum from the instrumental response using Eq. (3). The factors dependent on neutrino mass (Eq. (27)) are then subject to Abel transformation (Abel integral equation, Eq. (22)), the solution of which (Eq. (25)) is proven in to be a superposition of the step functions (Eq. (12)).

The Abel transformation was numerically tested on theoretically calculated beta spectra. It was shown that this method is sensitive down to a relative contribution of about $10^{-3}$ for heavy neutrinos and can set upper limits on the neutrino mass of 0.1 eV for $^3$H and 0.03 keV for $^{35}$S. We have also numerically tested two possible experimental deviations in beta spectra: shape correction and energy nonlinearities. It was shown that a smooth, moderate shape correction up to 1% at low energy changes the slope of the step function but does not affect the numerical recognition of either zero-mass or heavy neutrino. Likewise, heavy neutrino detection is not affected by energy nonlinearity. Nevertheless, the energy nonlinearity can simulate an artificial neutrino mass close to the endpoint of β spectrum. However, if the Abel method is applied to such a case, the step function is not sharp and, therefore, the hypothesis of neutrino mass can be rejected. If not neutrino mass, the proposed method can indicate some discrepancies between measured and calculated β spectra.

The math-theoretical methods derived here are inherently non-statistical, however they have possible limitations arising from numerical errors, theoretical considerations, and statistical fluctuations present in the observed spectra. They are discussed below in that order.

It was shown that at the $10^{-3}$ level, the Abel-transformation method resulted in relative fluctuations below 2%, originating from rounding errors in the calculations. Such rounding errors could be possibly lowered, resulting in a higher sensitivity, if higher-precision arithmetic were used in the numerical calculations.

The theory of beta decay with all possible corrections, from which theoretical beta spectra are derived, can be given only to a finite accuracy, which is a limiting factor for neutrino mass searches in beta decay, regardless of the method used [24].

Finally, the statistical fluctuations in the observed beta spectra limit the sensitivity of neutrino-mass searches in beta decay by any method, including this one. Owing to the fact that the observed spectrum is discrete, there are Poisson fluctuations in each spectrum channel. The variation coefficient of the Poisson fluctuations is given by $1/\sqrt{w}$, where $w$ is the number of observed counts in the channel [33]. As seen in Fig. 1, the fluctuations are expected to be lower at the low beta energy region and higher close to the endpoint. The fluctuations present in the observed spectrum will translate into the fluctuations of any solution derived from the spectrum. It is, therefore, imperative to have the observed statistics in the data as high as possible. One way to alleviate problems with low statistics could be fitting of a polynomial to such data and applying integral equations to the polynomial fit instead of the original data points.

## Acknowledgments

Thanks are due to C. J. Bradt for his valuable comments.

## References

[1]   E. Fermi, Tentativo di una teoria dell'emissione dei raggi «beta», Ric. Scientifica **4**(2) (1933) 491-495.
[2]   E. Fermi, Versuch einer theorie der β-strahlen, Z. Physik **88** (1934) 161-171.
[3]   F.L. Wilson, Fermi's theory of beta decay, Am. J. Phys. **36** (1968) 1150-1160.

[4] C. Giunti, C.W. Kim, Fundamentals of Neutrino Physics and Astrophysics (Oxford University Press, Oxford, 2007).

[5] O.G. Miranda, J.W.F. Valle, Neutrino oscillations and the seesaw origin of neutrino mass, Nucl. Phys. B **908** (2016) 436-455.

[6] G. Drexlin, V. Hannen, S. Mertens, C. Weinheimer, Current direct neutrino mass experiments, Adv. High En. Phys. 293986 (2013) 1-39.

[7] Planck Collaboration, Planck 2015 results XIII. Cosmological parameters, Astron. Astrophys. **594** A13 (2016) 1-63.

[8] GERDA Collaboration, Background-free search for neutrinoless double-β decay of $^{76}$Ge with GERDA, Nature **544** (2017) 47-52.

[9] H.V. Klapdor-Kleingrothaus, I.V. Krivosheina, A. Dietz, O. Chkvorets, Search for neutrinoless double beta decay with enriched $^{76}$Ge in Gran Sasso 1990–2003, Phys. Lett. B **586** (2004) 198-212.

[10] T.J. Loredo, D.Q. Lamb, Bayesian analysis of neutrinos observed from supernova SN 1987A, Phys. Rev. D **650** 63002 (2002) 1-39.

[11] C. Kraus, B. Bornschein, L. Bornschein *et al.*, Final results from phase II of the Mainz neutrino mass search in tritium *β* decay, Eur. Phys. J. C **40** (2005) 447-468.

[12] V.N. Aseev, A.I. Belesev, A.I. Berlev *et al.*, Upper limit on the electron antineutrino mass from the Troitsk experiment, Phys. Rev. D **84** 112003 (2011) 1-9.

[13] S. Mertens, Direct neutrino mass experiments, J. Phys.: Conf. Ser. **718** 022013 (2016) 1-9.

[14] F. Gatti, Microcalorimeter measurements, Nucl. Phys. B (Proc. Suppl.) **91** (2001) 293-296.

[15] M. Sisti, C. Arnaboldi, C. Brofferio *et al.*, New limits from the Milano neutrino mass experiment with thermal microcalorimeters, Nucl. Instr. Meth. Phys. Res. A **520** (2004) 125-131.

[16] P.T. Springer, C.L. Bennett, P.A. Baisden, Measurement of the neutrino mass using the inner bremsstrahlung emitted in the electron-capture decay of $^{163}$Ho, Phys. Rev. A **35** (1987) 679-689.

[17] S. Yasumi, H. Maezawa, K. Shima *et al.*, The mass of the electron neutrino from electron capture in $^{163}$Ho, Phys. Lett. B **334** (1994) 229-233.

[18] A. Hime, J.J. Simpson, Evidence of the 17-keV neutrino in the β spectrum of $^{3}$H, Phys. Rev. D **39** (1989) 1837-1850.

[19] E.B. Norman, B. Sur, K.T. Lesko, M.M. Hindi, R.-M. Larimer, T.R. Ho, J.T. Witort, P.N. Luke, W.L. Hansen, E.E. Haller, Evidence for the emission of a massive neutrino in the nuclear beta decay, J. Phys. G: Nucl. Part. Phys. **17** (1991) S291-S299.

[20] J.J. Simpson, A. Hime, Evidence of the 17-keV neutrino in the β spectrum of $^{35}$S, Phys. Rev. D **39** (1989) 1825-1836.

[21] A. Hime, N.A. Jelley, New evidence for the 17 keV neutrino, Phys. Lett. B **257** (1991) 441-449.

[22] A. Hime, Do scattering effects resolve the 17-keV conundrum? Phys. Lett. B **299** (1993) 165-173.

[23] F.E. Wietfeldt, E.B. Norman, The 17-keV neutrino, Phys. Rep. **273** (1996) 149-197.

[24] M. Bahran, W.R. Chen, G.R. Kalbfleisch, Fermi theory of nuclear decay and heavy neutrino searches, Phys. Rev. D **47** (1993) R759-R763.

[25] X. Mougeot, Systematic comparison of beta spectra calculations using improved analytical screening correction with experimental shape factors, Appl. Radiat. Isot. **109** (2016) 177-182.

[26] H. Behrens, J. Jänecke, Numerical Tables for Beta-Decay and Electron Capture, in: K.-H. Hellwege (Ed.), Landolt-Börnstein Numerical Data and Functional Relationships in Science and Technology (Springer-Verlag, Berlin, 1969).

[27] M.E. Rose, A note on the possible effect of screening in the theory of beta-disintegration, Phys. Rev. **49** (1936) 727-729.

[28] L. Durand, Electron screening corrections to beta-decay spectra, Phys. Rev. **135** (1964) B310-B313.

[29] J.M. Carmona, J.L. Cortés, Testing Lorentz invariance violations in the tritium beta-decay anomaly, Phys. Lett. B **494** (2000) 75-80.

[30] W.J. Huang, G. Audi, M. Wang, F.G. Kondev, S. Naimi, Xing Xu, The Ame2016 atomic mass evaluation, (I). Evaluation of input data; and adjustment procedures, Chin. Phys. C **41**(3) 030002 (2017) 1-344.

[31] D. Porter, D.S.G. Stirling, Integral Equations (Cambridge University Press, Cambridge, 1993) Sects. 1.2, 1.3.4, 9.2.

[32] N.L. Johnson, S. Kotz, N. Balakrishnan, Continuous Univariate Distributions, Vol. 1 (Wiley-Interscience Publication, New York, 1994) Sect. 12.3.

[33] N.L. Johnson, A.W. Kemp, S. Kotz, Univariate Discrete Distributions (Wiley-Interscience, Hoboken, 2005) Sects. 1.2.11, 4.4.

[34] P.A.M. Dirac, The Principles of Quantum Mechanics (Oxford University Press, Oxford, 1988) Sect. III.15.

[35] Numerical Algorithms Group Ltd, Oxford, UK.

[36] Wolfram Research, Champaign, IL, USA.

[37] H. Margenau, G.M. Murphy, The Mathematics of Physics and Chemistry (Krieger, Huntington, 1976) Sect. 8.5.

[38] E.C. Titchmarsh, Introduction to the Theory of Fourier Integrals (Oxford University Press, Oxford 1967) Sect. 11.14.

[39] I.S. Gradshteyn and I.M. Ryzhik, Table of Integrals, Series, and Products, 8th Ed., D. Zwillinger (Ed.) (Academic Press, Boston, 2015) Sect. 2.61.